\documentclass[oldversion]{aa}
\usepackage{amsmath,graphics,amssymb}

\usepackage{natbib}
\bibpunct{(}{)}{,}{a}{}{,}
\defcitealias{nltei}{Paper I}

\newcommand{\zav}[1]{\left(#1\right)}
\newcommand{\ms}{\ensuremath{\text{M}_{\odot}}}
\newcommand{\kms}{\ensuremath{\mathrm{km}\,\mathrm{s}^{-1}}}
\newcommand{\msr}{\ensuremath{\ms\,\text{year}^{-1}}}

\newcommand{\hzav}[1]{\left[#1\right]}

\newcommand\de{\text{d}}

\allowdisplaybreaks

\begin{document}

\title{CMF models of hot star winds}
\subtitle{I. Test of the Sobolev approximation in the case of pure line
transitions}

\author{J.  Krti\v{c}ka\inst{1} \and J. Kub\'at\inst{2}}
\authorrunning{J. Krti\v{c}ka and J. Kub\'at}

\institute{\'Ustav teoretick\'e fyziky a astrofyziky P\v{r}F MU,
            CZ-611 37 Brno, Czech Republic, \email{krticka@physics.muni.cz}
           \and
           Astronomick\'y \'ustav, Akademie v\v{e}d \v{C}esk\'e
           republiky, CZ-251 65 Ond\v{r}ejov, Czech Republic}

\date{Received 21 January 2010}

\abstract{We provide hot star wind models with radiative force calculated using
the solution of comoving frame (CMF) radiative transfer equation. The wind
models are calculated for the first stars, O stars, and the central stars of planetary
nebulae. We show that without line overlaps and with solely thermal line
broadening the pure Sobolev approximation provides a reliable estimate of the
radiative force even close to the wind sonic point. Consequently, models with
the Sobolev line force provide good approximations to solutions obtained with
non-Sobolev transfer. Taking line overlaps into account, the radiative force
becomes slightly lower, leading to a decrease in the wind mass-loss rate
by roughly 40\%. Below the sonic point, the CMF line force is significantly lower
than the Sobolev one. In the case of pure thermal broadening, this does not
influence the mass-loss rate, as the wind mass-loss rate is set in the
supersonic part of the wind. However, when additional line broadening is present
(e.g., the turbulent one) the region of low CMF line force may extend outwards
to the regions where the mass-loss rate is set. This results in a decrease in the
wind mass-loss rate. This effect can at least partly explain the low wind mass-loss
rates derived from some observational analyses of luminous O stars.

}

\maketitle

\section{Introduction}

One of the most important galactic populations consists of massive
stars, because these stars dominate the spectra of many galaxies and 
contribute significantly to the mass and momentum input into the interstellar matter.
Moreover, massive stars end their active lives in gigantic explosions such as supernovae
or even possibly as the progenitors of gamma-ray bursts \citep[see][] {wooo,yola},
producing huge amounts of heavier elements.

An important property of hot stars that significantly influences
their final stages is the stellar wind \citep[see, e.g.,][for reviews dedicated
to hot star winds]{owopo,kkpreh,pulvina}. However, in
stellar evolution calculations it is usually unnecessary to
know detailed wind properties,
but just the amount of mass expelled from the star per unit of time (mass-loss
rate) as a function of stellar parameters (e.g., mass, effective temperature, radius,
surface metallicity). However, for many hot stars we simply 
cannot estimate
their true mass-loss rate with the precision necessary to calculate
evolutionary models. The situation may be less problematic for luminous O
stars, for which relatively good agreement between theoretical predictions and
observational results seems to exist \citep[hereafter
\citetalias{nltei}]
{pahole,vikolamet,nltei}. 

However, the agreement between theoretically predicted mass-loss rates
and those derived from observations may be
an
illusion caused by
the
neglect of 
some physical effects in the wind, such as clumping
\citep{bourak,martclump}.
%
As a result, the true mass-loss rates of O stars may be a few times lower than the
standard wind theory predicts. This seems to be supported by the observations of
\cite{fuj} of weak wind line profiles of \ion{P}{v}. Last but not least, 
the unexpected occurrence of symmetrical X-ray line profiles seems to require
relatively low wind mass-loss rates \citep{kram}.

Possibly
unreliable estimates of hot star wind mass-loss rates are
also problematic because altough more realistic
evolutionary stellar models can be calculated, by including, e.g.,~rotation
and magnetic fields, 
%
the wind mass-loss rates remain uncertain. Ideally, all
observational indicators of mass-loss rate and theoretical models should find
and predict similar
mass-loss rates. From the observational point of view, more
detailed models of line formation in inhomogeneous media may be necessary
to obtain reliable line profiles, and consequently also estimate mass-loss
rates 
\citep{osporcar,chuchcar}.

From the theoretical point of view, disagreement between theory and
observations would imply that some of the assumptions used for the hot-star wind
modeling are inaccurate. Part of the disagreement may be caused by using
incorrect abundances \citep{spravnez}, although 
the reason for a
disagreement remains mainly unclear. A thorough inspection
of all assumptions involved in the modeling is therefore strongly needed. As a first step
in this direction, we studied the influence of X-rays on the wind structure
of hot stars. It seems that X-rays alone cannot entirely explain the
disagreement between theory and observations 
\citep{nlteiii}
as their
influence on wind mass-loss rates is small and they do not strongly affect the
ionization fraction of many important ions, especially that of \ion{P}{v}. On
the other hand, the modified ionization equilibrium may affect the X-ray line
formation \citep{lidataky,nlteiii}, and too ling cooling time in the
post-shock region \citep{cobecru,nlteiii} may cause the so-called "weak wind
problem" \citep{bourak,martin,linymarko}.

One of the most important approximations in the hot-star wind modeling is the
Sobolev approximation \citep{sobolevprvni,cassob}, which enables
us
to solve the
line radiation transfer analytically. Some studies confirm its applicability in
the supersonic part of smooth line-driven winds 
\citep{hamki,ppk,pulpren}.
However,
the applicability of the Sobolev approximation is questionable especially in the
regions close to the photosphere bacause of the existence of strong source function
gradients \citep{opsim}. On the other hand, some models avoid using the Sobolev
approximation and use only the comoving-frame (hereafter CMF) method of solving
the radiative transfer equation \citep[e.g.,][]{graham}.

We decided to test the applicability of the Sobolev approximation
and include the CMF solution of the radiative transfer equation in our wind
models.
%
In this
first paper of a 
series,
we describe our method, and study the
applicability of the Sobolev approximation 
using models
neglecting continuum 
opacity.


\section{Basic model assumptions}

The models used in this paper are based on the NLTE wind models of
\citet[hereafter \citetalias{nltei}]{nltei}. Here we summarise only their basic
features and describe the inclusion of CMF line force.

We assume a spherically symmetric stationary stellar wind. The excitation and
ionization state of the considered elements is derived from the statistical
equilibrium (NLTE) equations. Ionic models are either adopted from the TLUSTY
grid of model stellar atmospheres \citep{ostar2003,bstar2006} or are prepared by
us using the data from the Opacity and Iron Projects \citep{top,
opc5,top1,savej,topt,bumez,napra,zel0,zel6,zel2,zel1,zel5,zel4,zel3}.
As in \citetalias{nltei}, the
solution of the radiative transfer equation for NLTE equations is artificially
split into two parts, namely the radiative transfer in either the continuum 
or in lines. The solution to the radiative transfer equation in
continuum is based on the Feautrier method in the spherical coordinates
\citep{sphermod,dis}, and the line radiative transfer is solved in the Sobolev
approximation \citep{cassob,rybashumrem} neglecting continuum opacity and line
overlaps.

In contrast to our previous models, the radiative transfer in lines used for the
calculation of the radiative force is solved in the CMF (see Sect.~\ref{cmfkap})
neglecting the continuum opacity. 
The line radiative force is calculated directly from the true chemical composition,
NLTE ionization and excitation balance, and CMF flux using data from the VALD
database \citep{vald1,vald2}. We do not use the
line-strength distribution function parameterized by force multipliers $k$,
$\alpha$, and $\delta$.

The flux at the surface (used as the lower boundary condition for the radiative
transfer in the wind) is taken from the H-He spherically symmetric NLTE model
stellar atmospheres of \citet[and references therein]{kub}.

\section{CMF calculation of the radiative force}
\label{cmfkap}

The radiative force is calculated using the solution of the spherically
symmetric CMF radiative transfer equation \citep[Eq.~(14.99)]{mih}
\begin{multline}
\label{rpzint}
\mu \frac{\partial I(\nu,\mu,r)}{\partial r}+
\frac{1-\mu^2}{r}\frac{\partial I(\nu,\mu,r)}{\partial \mu}\\*-
\frac{\nu v_r}{cr}\hzav{1-\mu^2+\mu^2\frac{r}{v_r}\frac{\de v_r}{\de r}}
\frac{\partial I(\nu,\mu,r)}{\partial \nu}\\*=
\eta(\nu,r)-\chi(\nu,r)I(\nu,\mu,r),
\end{multline}
where $I(\nu,\mu,r)$ is the intensity seen by the observer moving with the
wind at the radial velocity $v_r$, $\nu$ is the frequency, $\mu=\cos\theta$,
$\theta$ is the direction between the given ray and the radial direction, and
$\eta(\nu,r)$ and $\chi(\nu,r)$ are the line emissivity and opacity,
respectively, given by
\begin{subequations}
\label{chieta}
\begin{align}
\eta(\nu,r)&=\frac{2h\nu^3}{c^2}\sum_{i,j}\frac{\pi e^2}{m_\text{e}c}
\frac{n_j}{g_j}g_if_{ij}\varphi_{ij}(\nu),\\
\chi(\nu,r)&=\sum_{i,j}\frac{\pi e^2}{m_\text{e}c}
\zav{\frac{n_i}{g_i}-\frac{n_j}{g_j}}g_if_{ij}\varphi_{ij}(\nu),
\end{align}
\end{subequations}
where $n_i$ and $n_j$ are the number densities of individual states with
statistical weights $g_i$ and $g_j$ corresponding to the line transition
$i\leftrightarrow j$ with oscillator strength $f_{ij}$ and the line-profile
$\varphi_{ij}(\nu)$, and $m_\text{e}$ is the electron mass. Assuming the line
profile to be Gaussian 
produced by thermal broadening only,
$\varphi_{ij}(\nu)$ is given by
\begin{equation}
\varphi_{ij}(\nu)=\frac{1}{\sqrt\pi\Delta\nu_{ij}}
\exp\hzav{\frac{\zav{\nu-\nu_{ij}}^2}{\Delta\nu_{ij}^2}},
\end{equation}
where $\nu_{ij}$ is the laboratory line frequency, and the line broadening is
given by
\begin{equation}
\label{deltanub}
\Delta\nu_{ij}=\frac{\nu_{ij}}{c}\sqrt{\frac{2kT}{m_a}},
\end{equation}
and $m_a$ is the mass of a given atom. 
The number densities of individual levels in Eqs.~\eqref{chieta} are
calculated from
statistical equilibrium (NLTE) equations.

Writing Eq.~\eqref{rpzint}, we neglected advection and aberration terms, which is
justifiable in non-relativistic flows \citep[see, e.g.,][]{korku}. We also note
that by neglecting the spatial derivatives of intensity in Eq.~\eqref{rpzint} we
obtain the Sobolev approximation \citep{castor}.

Following \cite{mikuh}, we rewrite Eq.~\eqref{rpzint} for
rays with an impact parameter $p$
\begin{multline}
\label{rpzintchar}
\pm\frac{\partial I^\pm(\nu,p,z)}{\partial z}-
\frac{\nu v_r}{cr}\hzav{1-\mu^2+\mu^2\frac{r}{v_r}\frac{\de v_r}{\de r}}
\frac{\partial I^\pm(\nu,p,z)}{\partial \nu}\\*=
\eta(\nu,r)-\chi(\nu,r)I^\pm(\nu,p,z),
\end{multline}
where $+$ and $-$ refers to radiation flowing toward and away from the observer,
respectively,
$r=(p^2+z^2)^{1/2}$, and $z$ is the distance along the ray.
We transform Eq.~\eqref{rpzintchar} using intensity-like and flux-like
variables
\begin{subequations}
\begin{align}
u(\nu,p,z)&=\frac{1}{2}\hzav{I^+(\nu,p,z)+I^-(\nu,p,z)},\\*
v(\nu,p,z)&=\frac{1}{2}\hzav{I^+(\nu,p,z)-I^-(\nu,p,z)},
\end{align}
\end{subequations}
to obtain a system of partial differential equations
\begin{subequations}
\label{rpzuv}
\begin{gather}
\label{rpzuv1}
\frac{1}{\chi(\nu,r)}\frac{\partial u(\nu,p,z)}{\partial z}-
\gamma(\nu,p,z)\frac{\partial v(\nu,p,z)}{\partial \nu}=-v(\nu,p,z),\\
\label{rpzuv2}
\frac{1}{\chi(\nu,r)}\frac{\partial v(\nu,p,z)}{\partial z}-
\gamma(\nu,p,z)\frac{\partial u(\nu,p,z)}{\partial \nu}=S(\nu,r)-u(\nu,p,z),
\end{gather}
\end{subequations}
where
\begin{align}
\gamma(\nu,p,z)&=\frac{\alpha(r)}{r\chi(\nu,r)}\hzav{1-\mu^2+\beta(r)\mu^2},\\
\alpha(r)&=\frac{\nu v_r}{c},\\
\beta(r)&=\frac{r}{v_r}\frac{\de v_r}{\de r},\\
S(\nu,r)&=\frac{\eta(\nu,r)}{\chi(\nu,r)}.
\end{align}

The system of equations in Eq.~\eqref{rpzuv} is solved numerically using the long
characteristic method of \citet{mikuh}, which we modified slightly for the
present purpose (see Appendix~\ref{apnumcmf}). As we are interested in the
calculation of the radiative force using the $v$ variable at a particular grid
point, in contrast to \citet{mikuh} we specify $v$ at grid points, and $u$ in the
middle between them.

In our numerical solution of Eq.~\eqref{rpzuv}, we use the same spatial grid as for
the solution of hydrodynamical equations. The spacing of the frequency grid is
$\Delta\nu_\text{D}=\nu\sqrt{\zav{2kT_\text{C}/m_\text{C}}}/\zav{cf_\text{D}}$,
where $T_\text{C}$ is the pre-specified expected minimum wind temperature,
$m_\text{C}$ is the atomic mass of artificial metallic atom, and $f_\text{D}$ is
the multiplicative factor (see below). The CMF radiative transfer equation is
solved only for selected frequencies from the frequency grid that lie close to
some line. The selection of frequencies is controlled by two integer numbers
$n_\text{D}$, and NCERV \citep[cf.][]{hilmi}. For each line, we select
frequencies that lie within $n_\text{D}$ line Doppler widths $\Delta\nu_{ij}$.
Redward of the center of each line, we select each NCERV frequency up
to the frequency corresponding to the Doppler shift for the wind terminal
velocity. The numerical test showed that a sufficiently precise value of the
radiative force can be derived for the value of parameters
$m_\text{C}=60\,m_\text{H}$, where $m_\text{H}$ is the mass of hydrogen atom,
$f_\text{D}=2$, $n_\text{D}=5$, NCERV=30, and typically
$T_\text{C}=10\,000-20\,000\,$K.

The radiative force is calculated as an integral
\begin{multline}
\label{fcmf}
f_\text{rad}^\text{CMF}=\frac{1}{c}\int_0^\infty \chi(\nu,r) F(\nu,r)\,\de\nu
\\*=\frac{4\pi}{c}\int_0^\infty \de\nu
\int_0^1\de\mu\,\mu \chi(\nu,r) v(\nu,p,z).
\end{multline}
As the calculation of the CMF radiative force is rather time-consuming, we do
not calculate $f_\text{rad}^\text{CMF}$ during each iteration of
hydrodynamical variables, but adopt a different approach. We calculate the
ratio of the CMF and Sobolev line forces
\begin{equation}
\label{slozityzlomek}
c^\text{CMF}=\frac{f_\text{rad}^\text{CMF}}{f_\text{rad}^\text{Sob}}.
\end{equation}
By the Sobolev line force
$f_\text{rad}^\text{Sob}$,
we mean here the force calculated
by assuming the
Sobolev approximation
for radiative transfer,
neglecting line overlaps and using 
true line opacities and
the emergent flux from the
underlying
stellar atmosphere \citepalias[Eq.~(25) therein]{nltei}.
Unless the base density is known with a precision better than about 30\%, we
calculate $c^\text{CMF}$ only when the estimate of the base density is changed,
and keep $c^\text{CMF}$ fixed during the subsequent iterations of the
hydrodynamical structure. When the base density is known with a higher
precision, we calculate $c^\text{CMF}$ after each change of the hydrodynamical
structure. Moreover, because we solve the hydrodynamical equations using the
Newton-Raphson method, we have to calculate the derivatives of
$f_\text{rad}^\text{CMF}$ with respect to individual hydrodynamical variables.
These derivatives are approximated using the derivatives of the Sobolev line
force $f_\text{rad}^\text{Sob}$ multiplied by $c^\text{CMF}$. The force term in
the critical point condition \citepalias[see][]{nltei} is also multiplied by
$c^\text{CMF}$.

We note that direct use of Eq.~\eqref{slozityzlomek} causes instability in the
model convergence. The reason for these convergence problems may be numerical,
but this behavior may also be connected with line-driven instability
\citep{ocr,felpulpal}. To avoid this
(since we are seeking stationary solution and not evolution with time)
we introduced a weak smoothing of
$c^\text{CMF}$,
\begin{equation}
\label{vyhlad}
\overline c^\text{CMF}_d=\frac{1}{4}\zav{2c^\text{CMF}_d+c^\text{CMF}_{d-1}+
c^\text{CMF}_{d+1}},
\end{equation}
where $c^\text{CMF}_d$ is the value of $c^\text{CMF}$ at a given grid point $d$
(as for $d-1$ and $d+1$) and we use $\overline c^\text{CMF}_d$ instead of
$c^\text{CMF}_d$ in the models.
Our numerical tests showed that the smoothing Eq.~\eqref{vyhlad} does not
significantly affect the resulting radiative force.

\begin{table}[tb]
\caption{Radius $R_*$, mass $M$, and the effective temperature $T_\text{eff}$
of studied model stars}
\label{hvezpar}
\begin{center}
\begin{tabular}{ccccc}
\hline
\hline
\multicolumn{2}{c}{Star (model)} & $R_*$ & $M$ & $T_\text{eff}$\\
& & $[\text{R}_\odot]$ & $[\text{M}_\odot$] & [K]\\
\hline
\multicolumn{5}{c}{First stars}\\
\hline
\multicolumn{2}{c}{M500-1} & 11.1 &  50 & 50\,000 \\
\multicolumn{2}{c}{M500-2} & 33.7 &  50 & 29\,900 \\
\multicolumn{2}{c}{M500-3} & 72.0 &  50 & 20\,600 \\
\multicolumn{2}{c}{M500-4} & 303  &  50 & 10\,100 \\
\hline
\multicolumn{5}{c}{O stars}\\
\hline
$\xi$ Per      &  HD $24912$ & $14.0$ & $36$ & $35\,000$ \\
$\iota$ Ori    &  HD $37043$ & $21.6$ & $41$ & $31\,400$ \\
15 Mon         &  HD $47839$ &  $9.9$ & $32$ & $37\,500$ \\
               &  HD $54662$ & $11.9$ & $38$ & $38\,600$ \\
               &  HD $93204$ & $11.9$ & $41$ & $40\,000$ \\
$\zeta$ Oph    & HD $149757$ &  $8.9$ & $21$ & $32\,000$ \\
68 Cyg         & HD $203064$ & $15.7$ & $38$ & $34\,500$ \\
19 Cep         & HD $209975$ & $22.9$ & $47$ & $32\,000$ \\
\hline
\multicolumn{5}{c}{Central stars of planetary nebulae}\\
\hline
\multicolumn{2}{c}{NGC 2392}  & 1.5 &0.41 &40\,000 \\
\multicolumn{2}{c}{NGC 3242}  & 0.3 &0.53 &75\,000 \\
\multicolumn{2}{c}{IC 4637}   & 0.8 &0.87 &55\,000 \\
\multicolumn{2}{c}{IC 4593}   & 2.2 &1.11 &40\,000 \\
\multicolumn{2}{c}{He 2-108}  & 2.7 &1.33 &39\,000 \\
\multicolumn{2}{c}{IC 418}    & 2.7 &1.33 &39\,000 \\
\multicolumn{2}{c}{Tc 1}      & 3.0 &1.37 &35\,000 \\
\multicolumn{2}{c}{NGC 6826}  & 2.2 &1.40 &44\,000 \\
\hline
\end{tabular}
\end{center}
\end{table}

\section{Studied model stars}

In our study, we selected three types of stars to study more carefully the Sobolev
approximation in different wind environments (see Table~\ref{hvezpar}).

The stellar parameters of
the
first stars were obtained according to an evolutionary
calculation of initially zero-metallicity star with initial mass
$50\,\text{M}_\odot$ derived by \citet{bezmari}. For these models, we assumed a
stellar wind driven purely by CNO elements (which appear on the stellar surface
due to mixing) with a mass-fraction of CNO $Z=10^{-3}$.

The stellar parameters (effective temperatures and radii) of an O star sample were
derived using the model atmospheres with line blanketing 
\citep{rep,upice,martclump}.
Stellar masses were obtained using
evolutionary tracks either by ourselves (using \citealt{salek} tracks) or by
\citet{martclump}. For these stars, we assumed a solar chemical composition
\citep{asgres}.

The stellar parameters of central stars of planetary nebulae were taken from
\citet{btpau}, who derived them from UV spectroscopy. Helium abundance was
adopted from \citet{btkud}, for other elements we assumed a solar chemical
composition \citep[after][]{asgres}, which was for some stars slightly modified
according to \citet{btpau}.

\section{Comparison of CMF and Sobolev wind models}

\begin{figure}
\centering
\resizebox{0.8\hsize}{!}{\includegraphics{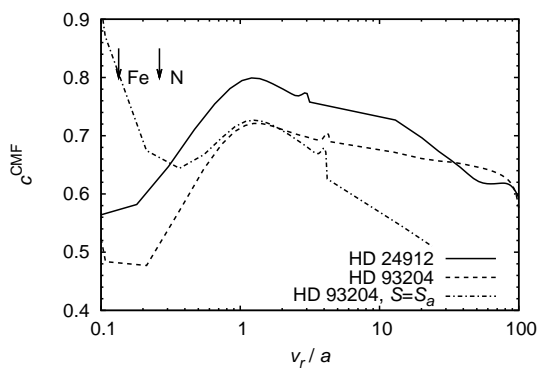}}
\caption{The ratio of the CMF to Sobolev line forces given by
Eq.~\eqref{slozityzlomek}
as a function of the wind
velocity plotted in the terms of the sound speed
for two selected stars. 
The 
dashed-dotted
line denotes a model with constant source function and constant
level populations equal to their values at the sonic point.
Arrows indicate the thermal speed of selected ions.}
\label{ccmf}
\end{figure}

We calculated wind models with both CMF and Sobolev line forces and compared the
final wind structure.
The resulting ratio of the CMF to Sobolev line forces $c^\text{CMF}$ for
selected stars is
shown
in Fig.~\ref{ccmf}.
We note that the Sobolev force was calculated using the flux from the stellar
atmosphere and by neglecting line overlaps.

For very 
low
wind velocities
$v_r\lesssim0.1a$ 
(where $a^2=2kT/m_\text{H}$),
the CMF force is large, $c^\text{CMF}>1$. This is most likely
partly connected with the boundary conditions, which are not completely
compatible with the wind \citep[cf.,][]{nory}.

\begin{figure}
\centering
\resizebox{0.8\hsize}{!}{\includegraphics{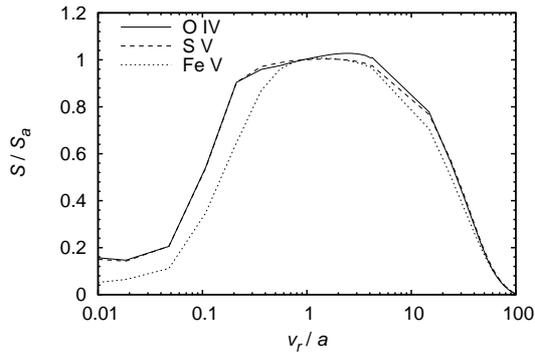}}
\caption{The line source function for resonance lines of \ion{O}{iv} at
790\,\AA, \ion{S}{v} at 786\,\AA, and \ion{Fe}{v} at 388\,\AA\ as a function of
relative wind velocity. The source function is plotted relative to its value at
the sonic point $S_a$ for the wind model of HD~93204.}
\label{zdroj}
\end{figure}

%
For velocities of about
one tenth of the sound speed, there is an apparent minimum of $c^\text{CMF}$. In
some cases, the ratio $c^\text{CMF}$ could even be negative, which corresponds
to a negative CMF
radiative force. The Sobolev approximation is not applicable in this region, but
a low value of the radiative force is also connected with positive
source function gradients. For subsonic velocities, the Doppler shift is less
important, and the line radiative transfer is given basically by the static
radiative transfer equation. In the optically thick regions, for
frequencies corresponding to line transitions 
%
it follows from Eq.~\eqref{rpzuv2}
$u\approx S$, from Eq.~\eqref{rpzuv1} $v\approx-(1/\chi)\de S/\de z$, and the
radiative force is proportional to the negative of the derivative of the source function
$f_\text{rad}^\text{CMF}\sim-\de S/\de z$ (see~Eq.~\eqref{fcmf}, and
\citealt{nory}).
Because the line source function increases here 
(see Fig.~\ref{zdroj}),
the line radiative force at
low
velocities may even be negative.
For a constant source function, the minimum of $c^\text{CMF}$ close to the
star is significantly weaker (see Fig.~\ref{ccmf}).
The source function minimum
below
the sonic
point is caused by a local temperature minimum, because the line
source function of optically thick lines
(which are, consequently, in detailed radiative balance)
close to the star
$S\approx n_j/n_1\sim (n_j/n_1)^*$
(asterisk denotes LTE value)
depends on temperature.
Another source function minimum for non-Sobolev source function due to velocity
field curvature was also found by \citet{selma}, and \citet{opsim}.
We note that in the case of the resonance lines plotted in Fig.~\ref{zdroj} the line source
function at larger radii is roughly proportional to $S\sim n_j/
n_1
\sim r^{-3}$
\citep[e.g.,][]{kupul}.

The minimum of $c^\text{CMF}$
close to the star
is also connected with the 
velocity gradient changing significantly
within the resonance zone. Thus, a given line also picks up the
radiation corresponding to a lower velocity gradient leading to a further reduction
in the radiative force. For velocities comparable to or 
higher
than the ion
thermal speed, the lines are deshadowed because of the Doppler effect, and the
radiative force increases. We note that we only include the thermal broadening,
hence these effects occur for velocities 
lower
than the sound
speed.

As the wind accelerates, the ratio of the CMF to Sobolev line force increases
and
reaches a value
close to one for velocities 
higher
than the thermal speed of the wind
driving ions (Fig.~\ref{ccmf}). This is unsurprising, because the Sobolev
approximation is applicable to regions with a large velocity gradient, which
exist
already close to the sonic point
$v_r=a$.
%
Owing to line overlaps, $c^\text{CMF}$
is
less than
one in the outer wind regions, where it reaches only $0.7-0.8$.

\begin{figure}
\centering
\resizebox{0.8\hsize}{!}{\includegraphics{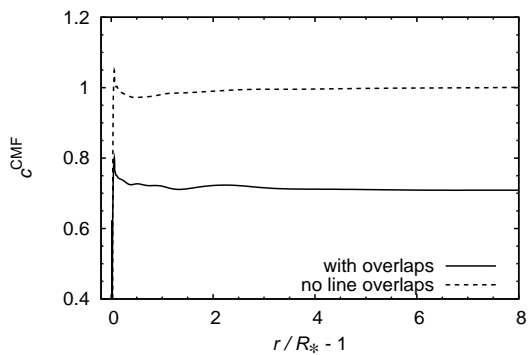}}
\caption{The radial variation 
in
the ratio of the CMF to Sobolev line forces in
the wind model of Tc 1 central star with and without line overlaps.}
\label{ccmf2}
\end{figure}

To test the influence of line overlaps, we calculated the radiative force with
only 50 carefully selected optically thick lines that do not overlap (see
Fig.~\ref{ccmf2}). The pronounced minimum for velocities 
lower
than the sound
speed is still present here, 
but
in the outer regions the value of
$c^\text{CMF}$ is approximately one, supporting the validity of the
Sobolev approximation for supersonic velocities.

\begin{figure}
\centering
\resizebox{0.8\hsize}{!}{\includegraphics{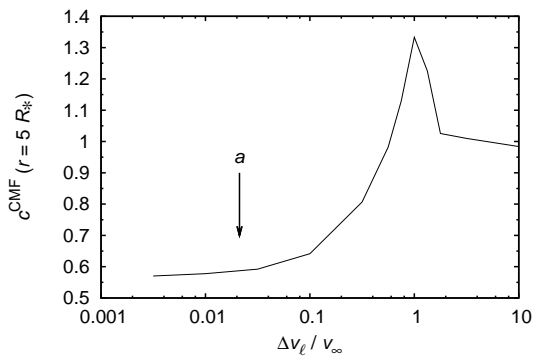}}
\caption{The ratio of the CMF to Sobolev line forces at radius $r=5\,R_*$
in the Tc1 wind model
in the dependence on the line shift. The value of hydrogen thermal speed is
denoted in the figure.}
\label{tc1nepdl}
\end{figure}

To understand more clearly the influence of line overlaps on the radiative
force,
we constructed
another
artificial
line list using
our
set of non-overlapping
lines. Each line in this new line list is counted twice with all parameters
being
completely the same, however with a line center shifted by $\nu_{ij}\Delta
v_\ell/
c$,
where
$\Delta v_\ell$
is a free parameter. For $\Delta v_\ell\ll a$,
all twin lines completely overlap leading to a significant decrease in the
radiative force with respect to the Sobolev one that does not account for the
line overlaps (see Fig.~\ref{tc1nepdl}). For $\Delta v_\ell>a$, the lines at a
given point do not overlap, but one of the twin lines "sees" the flux
absorbed by the second line, leading to a reduction in the radiative force even
in this case. For $\Delta v_\ell\approx v_\infty$, one of the lines is affected
by the
emission from the second one, leading to
an increase in the CMF radiative force relative to the Sobolev one.

A similar reduction in the line force by multiline effects
was found by \citet{pulpren}. We note that the multiline effects were also studied 
with respect to the multiple radiative momentum deposition in Wolf-Rayet star
winds \citep{gamom}. However, these effects are probably of minor importance
here due to the low density of the studied winds.

\begin{table}[t]
\caption{Comparison of calculated wind parameters derived using CMF and Sobolev
line forces}
\label{cmfsob}
\begin{center}
\begin{tabular}{lc@{\hspace{1mm}}rc@{\hspace{1mm}}r}
\hline
\hline
Star  & $\dot M^\text{Sob}$ & $v_\infty^\text{Sob}$ & $\dot M^\text{CMF}$
& $v_\infty^\text{CMF}$ \\
& [\msr] & [\kms]& [\msr] & [\kms]\\
\hline
\multicolumn{5}{c}{First stars}\\
\hline
M500-1 & $6.3\times10^{-08}$ & $2750$ & $6.1\times10^{-08}$ & $4010$ \\
M500-2 & $4.0\times10^{-07}$ & $1930$ & $2.0\times10^{-07}$ & $1310$ \\
M500-3 & $2.1\times10^{-07}$ & $580$ & $1.4\times10^{-07}$ & $790$ \\
M500-4 & $3.8\times10^{-08}$ & $600$ & $2.8\times10^{-08}$ & $620$ \\
\hline
\multicolumn{5}{c}{O stars}\\
\hline
  HD $24912$ & $4.4\times10^{-7}$ & $2270$ & $2.3\times10^{-7}$ & $2030$\\
  HD $37043$ & $6.2\times10^{-7}$ & $2340$ & $4.1\times10^{-7}$ & $2000$\\
  HD $47839$ & $2.2\times10^{-7}$ & $3080$ & $1.0\times10^{-7}$ & $2970$\\
  HD $54662$ & $7.9\times10^{-7}$ & $2190$ & $4.1\times10^{-7}$ & $2050$\\
  HD $93204$ & $1.3\times10^{-6}$ & $2290$ & $5.9\times10^{-7}$ & $2080$\\
 HD $149757$ & $4.7\times10^{-8}$ & $2040$ & $2.9\times10^{-8}$ & $1860$\\
 HD $203064$ & $5.7\times10^{-7}$ & $2080$ & $3.8\times10^{-7}$ & $1780$\\
 HD $209975$ & $8.4\times10^{-7}$ & $2430$ & $5.5\times10^{-7}$ & $1960$\\
\hline
\multicolumn{5}{c}{Central stars of planetary nebulae}\\
\hline
NGC 2392  & $3.7\times10^{-8}$ & $490$  & $1.8\times10^{-8}$ & $500$\\
NGC 3242  & $3.1\times10^{-9}$ & $2000$ & $2.0\times10^{-9}$ & $1890$ \\
IC 4637   & $3.1\times10^{-8}$ & $1440$ & $1.4\times10^{-8}$ & $1270$ \\
IC 4593   & $7.4\times10^{-8}$ & $730$  & $3.8\times10^{-8}$ & $660$\\
He 2-108  & $9.5\times10^{-8}$ & $730$  & $4.7\times10^{-8}$ & $700$ \\
IC 418    & $9.5\times10^{-8}$ & $730$  & $4.7\times10^{-8}$ & $700$ \\
Tc 1      & $2.8\times10^{-8}$ & $870$  & $1.8\times10^{-8}$ & $800$ \\
NGC 6826  & $1.8\times10^{-7}$ & $870$ & $7.6\times10^{-8}$ & $790$ \\
\hline
\end{tabular}
\end{center}
\end{table}

The CMF radiative force, which is lower than the Sobolev one because of
line overlaps,
causes a decrease in the mass-loss rate of CMF models with respect
to Sobolev ones (see Table~\ref{cmfsob}). The ratio of CMF to Sobolev mass-loss
rates is about $0.58$.
The only exception is the model M500-1, for which the CMF mass-loss rate is
nearly the same as the Sobolev one. The reason is that the star is so hot, that
the wind is accelerated mainly by a
dozen
\ion{O}{v} and \ion{O}{vi} lines.
For a critical point velocity, these lines do not
overlap, hence $c^\text{CMF}\approx1$, and the CMF and Sobolev
mass-loss rates are nearly the same.

\begin{figure}
\centering
\resizebox{0.8\hsize}{!}{\includegraphics{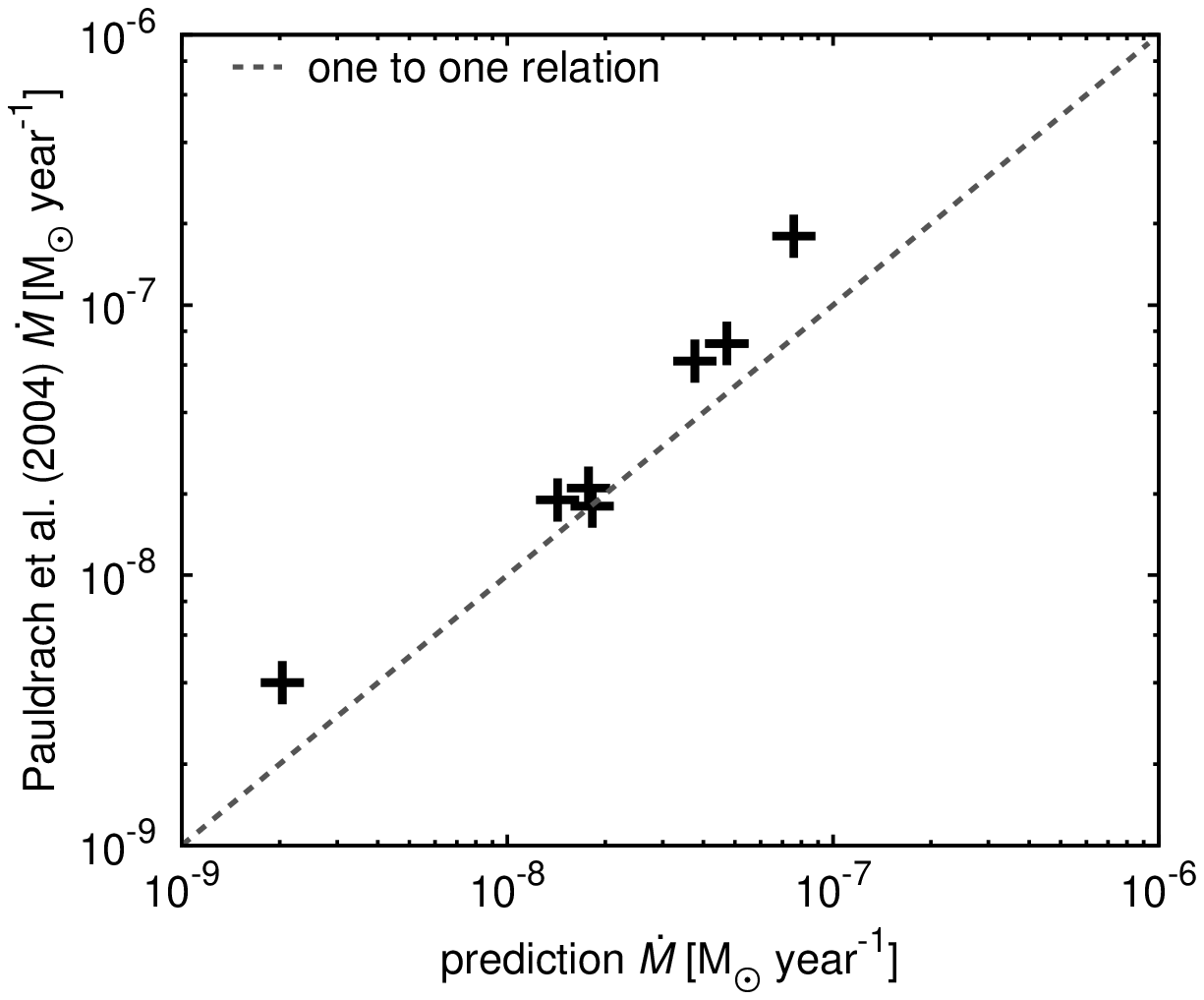}}
\resizebox{0.8\hsize}{!}{\includegraphics{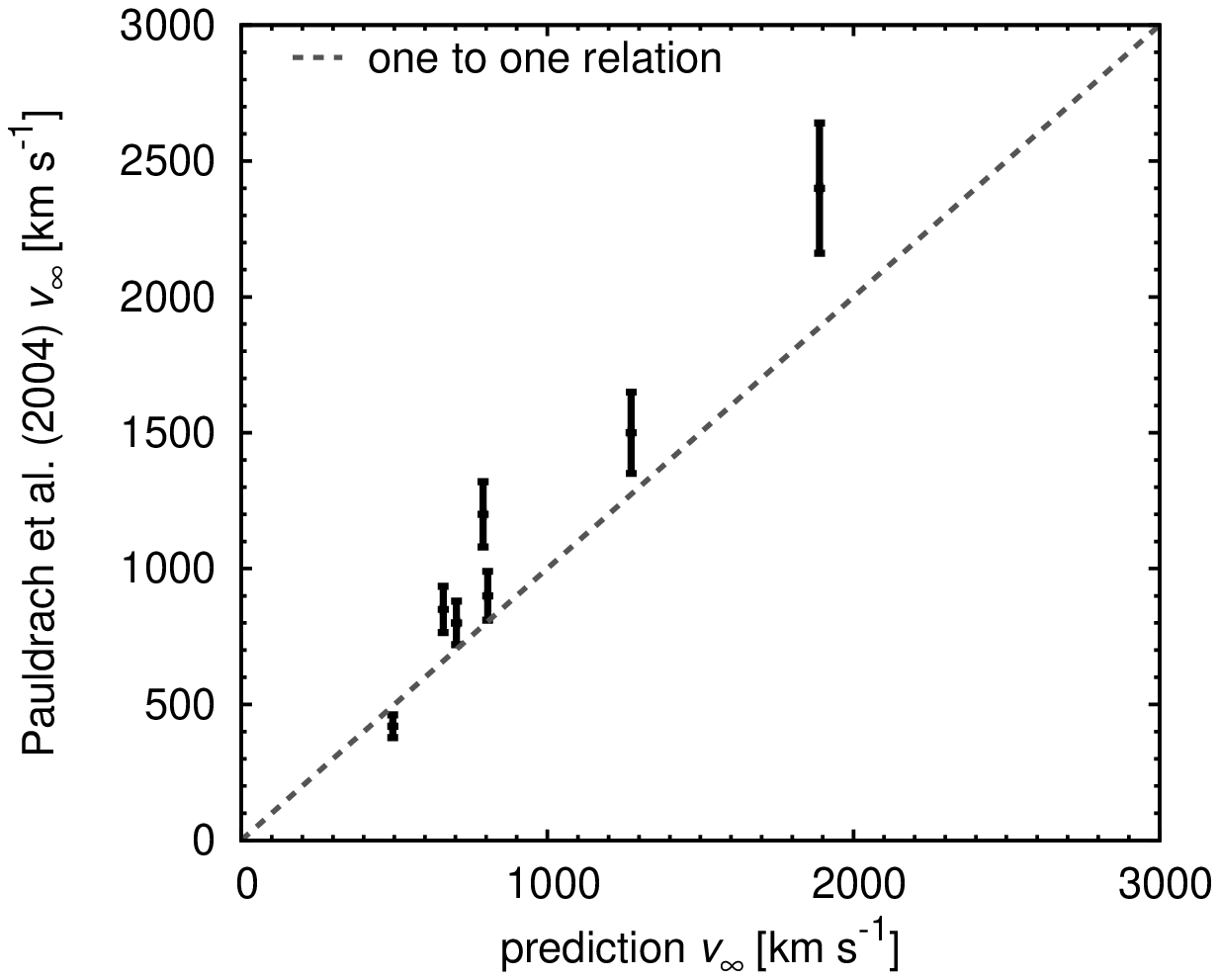}}
\caption{Comparison of our derived mass-loss rates (upper panel) and terminal
velocities (lower panel) of the central stars of planetary nebulae
with those derived by \citet{btpau}.}
\label{srpbtpau}
\end{figure}

The resulting wind parameters of the central stars of planetary nebulae can be compared
with those derived from observations by \citet[see Fig.~\ref{srpbtpau}]{btpau}.
There is reasonable agreement between the wind parameters predicted by ourselves and
those derived by \citet[see Fig.~\ref{srpbtpau}]{btpau}. The mass-loss rates of
\citet{btpau} are on average a factor of about $1.6$ higher than those
derived by ourselves. This is most likely partly because of the simplifications
included in 
our
code, e.g., the neglect of continuum opacity sources, and partly by the different
abundances adopted.

\section{Models with base turbulence}

\begin{figure}
\centering
\resizebox{0.8\hsize}{!}{\includegraphics{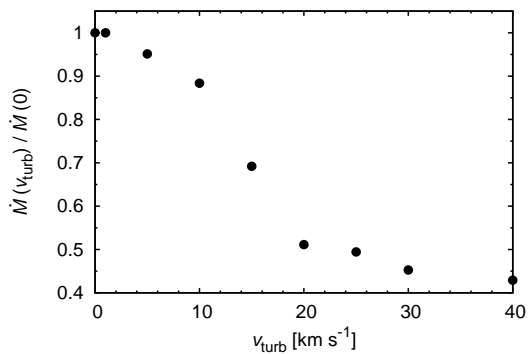}}
\caption{The mass-loss rate of HD~209975 in the models with additional
turbulent line broadening relative to the models with zero turbulent
velocity.}
\label{turbo}
\end{figure}

The existence of a region close to the stellar surface where the CMF line force is
low compared to the Sobolev one (see Fig.~\ref{ccmf}) is partly caused by the
source function gradients at the wind base and partly by the 
Sobolev approximation not being applicable to the subsonic regions. The CMF line
force increases at the moment when the line starts to absorb the radiation that
has not been absorbed yet, i.e., the radiation from the line wing. Because up to now
we have assumed pure thermal line broadening, the velocity width of 
low
CMF line
force is of the order of the metallic thermal speed (which is roughly $0.13a$ in the
case of iron). The wind mass-loss rate in our models is determined 
in the region of
supersonic wind, close to the critical point where the wind velocity approaches
the speed of radiative-acoustic waves \citep{abb,fero}. Thus, the
region of low CMF line force close to the star does not significantly affect the
wind mass-loss rate.

However, if the line broadening were larger (due to surface turbulence), then
the region of 
low
CMF line force could spread out to large velocities
comparable to the turbulent one. When the turbulent velocity is
comparable to the critical point velocity, below which the wind mass-loss rate
is set, this could cause a significant decrease in the wind mass-loss rate.
To test this, we calculated wind models with additional line broadening, which we
attributed to the turbulent one. In this case, the line profile width is given
not by Eq.~\eqref{deltanub}, but by
\begin{equation}
\label{deltanubtur}
\Delta\nu_{ij}=\frac{\nu_{ij}}{c}\sqrt{v_\text{turb}^2+\frac{2kT}{m_a}},
\end{equation}
where $v_\text{turb}$ is the adopted turbulent velocity.

The results of numerical models indicate that with increasing turbulent broadening
the velocity width of 
low
CMF line force increases leading to a lower mass-loss
rate (see Fig.~\ref{turbo}, cf.~\citealt{lucyinthewind}). Hence, in the
presence of turbulence the wind parameters may not depend only on the basic
stellar parameters (effective temperature, radius, mass) but also on the line
turbulent broadening. Moreover, this effect can possibly 
be one of the reasons
why the
mass-loss rates derived from observational analyses that take the clumping into
account \citep{bourak,martclump} are systematically lower than the predicted
ones.

For velocities 
higher
than a few times the turbulent one, the Sobolev approximation
should be applicable. At these high velocities, one expects that the line
force becomes close to the Sobolev one. Because now the same force (as in the
model with zero turbulent broadening) accelerates the wind of lower density, one
expects the terminal velocity to increase
\citep[e.g.,][]{gagra},
becoming much 
higher
than the observed one. However, our models do not predict a significant increase
in
the terminal
velocity $v_\infty$,
which is in the range $
%
1900-2200\,\kms$ for the
wind models of HD~209975
with different turbulent broadening.
This is caused by the stronger blocking of
stellar radiation by increased line overlaps mainly in the region with
$v_r\lesssim a$. 
We note that
lines broadened by turbulent motions
are able to block the flux more
efficiently than
lines broadened purely thermally.
%

Observational studies
consider
the turbulence already present in the photospheres of O stars
\citep[e.g.,][]{bourak,boula,martin,martclump} with turbulent velocities
of about
$2-25\,\kms$.
Macroturbulent
velocities in B supergiants
may be
even
higher,
about $30-100\,\kms$ \citep{howturb,kuraci}.
Convective layers and surface pulsational 
motions are
also expected theoretically \citep{cant,ae}.
Turbulence can spread
in the wind \citep{felpulpal}, leading to a decrease in the wind
mass-loss rate, as shown here.
We also note 
that many O stars exhibit turbulent velocities in the range $10-20\,\kms$,
where we expect a high sensitivity of the predicted mass-loss rate to the
turbulent velocity (see Fig.~\ref{turbo}).

The
basic
results presented here will, in the future, be tested
in more detail
using models that also account for the continuum opacity and CMF line
source function
in a separate study.
%

\section{The solution topology}

\begin{figure}
\centering
\resizebox{0.8\hsize}{!}{\includegraphics{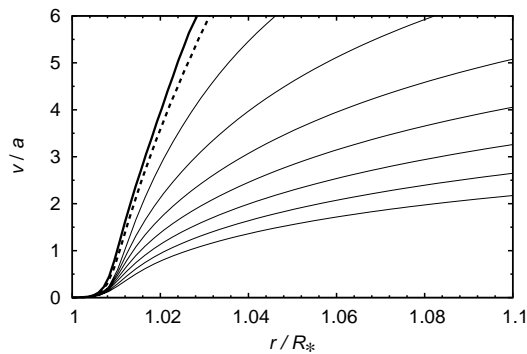}}
\caption{The dependence of the radial velocity on radius for solutions with
different boundary densities (mass-loss rates) close to the stellar surface for
the model star 500-1. Each consecutive model (from down to up) differs by a
factor of $1.5$ in the mass-loss rate (thin lines). The thick line denotes the
unique solution that smoothly intercepts the critical point. The dashed line
denotes solution with the Sobolev line force.}
\label{topca}
\end{figure}

In Fig.~\ref{topca}, we plot solutions with different base densities (mass-loss
rates). In generall, with increasing base density the wind velocity increases
until the density reaches a maximum value. There is no solution that is smooth
out to large radii for the densities 
higher
than the maximum one. The solution
with maximum density is very similar to the critical solution of Sobolev models
(see Fig.~\ref{topca}). Moreover, there are many solutions that smoothly pass
through the sonic point $v=a$ for different mass-loss rates. 

This indicates that the critical point of non-Sobolev models is close to the CAK
critical point \citep{cak} and that the sonic point is {\em not} a point where
the wind mass-loss rate is determined. The reason is that even in the
non-Sobolev models the radiative force is not given locally by wind density and
velocity, but depends on the wind properties in a close neighborhood of a
studied point. This dependence on the non-local properties at its limit
approaches the Sobolev approximation for very thin resonance layers (for very
large velocity gradients). 

\section{Conclusions}

We have presented hot star wind models in which the radiative force is calculated using the
solution of the comoving frame (CMF) radiative transfer equation. The wind models
were calculated for three different groups of stellar parameters (corresponding
to evolved first stars, O stars, and the central stars of planetary nebulae) to
compare the CMF and Sobolev radiative forces
for a broader range of stellar parameters.

The comparison of the CMF radiative force with an approximate one calculated
by assuming the Sobolev approximation showed that the Sobolev line force is slightly
higher
due to the neglect of line overlaps. Thus, the mass-loss rate of
wind models that include the Sobolev line force is on average a factor of about $1.7$
higher
than a more realistic one calculated using CMF wind models. However, we note
that the
simple
Sobolev approximation
applied here for reference
can be improved to account for line overlaps and
continuum absorption \citep{olson,humrsrybou,sobdif,sobreci}. We emphasize that
modern hot star wind models include line overlaps
\citep[e.g.,][]{vikolamet,pahole,graham}.

Without line overlaps in the case of purely thermal line broadening, the Sobolev
approximation provides a reliable estimate of the radiative force even close to
the wind sonic point. The CAK model therefore provides a good approximation for
a solution obtained with non-Sobolev transfer. Below the sonic point, the CMF line
force is significantly lower than the Sobolev one partly because of the strong gradients
in the source function. This does not influence the mass-loss rate, as the wind
mass-loss rate is set in the supersonic part of the wind below the critical
point. However, when additional line broadening is present (e.g., the turbulent
one) then the region of low CMF line force may extend outwards to the regions
where the mass-loss rate is set. This results in a significant decrease in the
wind mass-loss rate. We note that this is not a shortcoming of the Sobolev
approximation because the Sobolev approximation is applicable to velocities
higher
than the turbulent velocity in this case.

The influence of turbulent line broadening may cause a dependence of the
wind mass-loss rate on the atmospheric turbulent motions. Because
theoretical models are
not
yet
able to predict the atmospheric turbulent motions in hot stars in detail, we
are unable to provide reliable wind mass-loss rate predictions until the
theory of the atmospheric turbulence develops considerably \citep[however
see][]{cant,ae}. Nowadays, hot star evolution seems to be a deterministic one
depending only on the initial stellar parameters, i.e., mass, metallicity, and
rotational rate. However, as the properties of atmospheric turbulent motions
seem to be a non-trivial function of stellar parameters, the evolution of hot
stars may become less deterministic, becoming instead dependent on free parameters
describing the role of surface turbulence.

\begin{acknowledgements}
This work was supported by grant GA \v{C}R 205/07/0031. The Astronomical
Institute Ond\v{r}ejov is supported by the project AV0\,Z10030501.
\end{acknowledgements}

\newcommand{\actob}{in Active OB-Stars:
        Laboratories for Stellar \& Circumstellar Physics,
        ed. S. \v{S}tefl, S. P. Owocki, \& A.~T.
        Okazaki (ASP, San Francisco)}

\onecolumn
\appendix

\section{The solution of CMF radiative transfer equation}
\label{apnumcmf}

To calculate the radiative force, the \citet{mikuh} method for the
solution of the CMF radiative transfer equation is modified in such a way that
the $v$ variable is specified on the spatial grid, and $u$ on the intermediate
one.

Following the notation of \citet{mikuh}, the depth index $d$ increases inward,
$r_1=R_\text{out}>r_2>\dots>r_\text{ND}=R_*$, where $R_\text{out}$ is the radius
of the outer model boundary. The impact parameters $p$ are labeled in order of
increasing size by index $j$, $p_1=0<p_2<\dots<p_\text{NC}
<\dots<p_{\text{NC}+\text{ND}}$, where NC is the number of rays intersecting the
core. Along each ray with impact parameter $p_j$, we define grid in $z$ and
optical depth $\tau$, $z_{j1}=\zav{R_\text{out}^2-p_j^2}^{1/2}>
z_{j2}=\zav{r_2^2-p_j^2}^{1/2}>\dots>z_{j,\text{NI}_j}$, where
$\text{NI}_j=\text{ND}$ for $j\leq\text{NC}$, and
$\text{NI}_j=\text{ND}+\text{NC}+1-j$ for $\text{NC}<j\leq\text{ND}+\text{NC}$,
$\tau_{j1}=0<\tau_{j2}<\dots<\tau_{j,\text{NI}_j}$. The frequencies are labeled
by index $k$ in order of decreasing values, $\nu_1>\nu_2>\dots>\nu_\text{NF}$.

We assume that $v$ is specified on the depth grid, and $u$ is specified at
intermediate grid points labeled by $d\pm\frac{1}{2}$. Suppressing the ray index
$j$ in the following, we define on each ray $p_j$
\begin{align}
\chi_{k,d+1/2}&=\frac{1}{2}\hzav{\chi(\nu_k,z_{d+1})+\chi(\nu_k,z_{d})},\\
\Delta\tau_{k,d+1/2}&=\chi_{k,d+1/2}\zav{z_d-z_{d+1}},\\
\Delta\tau_{k,d}&=\frac{1}{2}\zav{\Delta\tau_{k,d+1/2}+\Delta\tau_{k,d-1/2}}.
\end{align}
The difference form of the system of the equations in Eq.~\eqref{rpzuv} is
\begin{subequations}
\begin{gather}
\label{rpzdifv}
\frac{u(\nu_k,z_{d+1/2})-u(\nu_k,z_{d-1/2})}{\Delta\tau_{k,d}}=v(\nu_k,z_{d})+
\frac{\gamma_{k,d}}{\Delta\nu_{k-1/2}}\hzav{v(\nu_k,z_{d})-v(\nu_{k-1},z_{d})},\\
\label{rpzdifu}
\frac{v(\nu_k,z_{d+1})-v(\nu_k,z_{d})}{\Delta\tau_{k,d+1/2}}=u(\nu_k,z_{d+1/2})-
S(\nu_k,z_{d+1/2})+\frac{\gamma_{k,d+1/2}}{\Delta\nu_{k-1/2}}
\hzav{u(\nu_k,z_{d+1/2})-u(\nu_{k-1},z_{d+1/2})},
\end{gather}
\end{subequations}
where $d=2,\dots,\text{NI}-1$, and
\begin{align}
\Delta\nu_{k-1/2}&=\nu_{k-1}-\nu_k,\\
\gamma_{k,d+1/2}&=\frac{\alpha_{d+1/2}}{r_{d+1/2}\chi_{k,d+1/2}}
\zav{1-\mu_{d+1/2}^2+\beta_{d+1/2}\mu_{d+1/2}^2},\\
\gamma_{k,d}&=\frac{\alpha_{d}}{r_{d}\chi_{k,d}}
\zav{1-\mu_{d}^2+\beta_{d}\mu_{d}^2}.
\end{align}
Solving Eq.~\eqref{rpzdifu} for $u(\nu_k,z_{d+1/2})$, we obtain
\begin{equation}
\label{ures}
u(\nu_k,z_{d+1/2})=\frac{v(\nu_k,z_{d+1})-v(\nu_k,z_{d})}
  {(1+\delta_{k-1/2,d+1/2})\Delta\tau_{k,d+1/2}}+
\frac{\delta_{k-1/2,d+1/2}}{1+\delta_{k-1/2,d+1/2}}u(\nu_{k-1},z_{d+1/2})+
\frac{S(\nu_k,z_{d+1/2})}{1+\delta_{k-1/2,d+1/2}},
\end{equation}
where
\begin{equation}
\delta_{k-1/2,d+1/2}=\frac{\gamma_{k,d+1/2}}{\Delta\nu_{k-1/2}}.
\end{equation}
Substituting Eq.~\eqref{ures} into Eq.~\eqref{rpzdifv}, we derive a linear system
of equations for $v(\nu_k,z_{d})$
\begin{multline}
\label{rpzdif}
\frac{1}{\Delta\tau_{k,d}}
\left[\frac{v(\nu_k,z_{d+1})}
      {\Delta\tau_{k,d+1/2}(1+\delta_{k-1/2,d+1/2})}-
      v(\nu_k,z_{d})\zav{\frac{1}{\Delta\tau_{k,d+1/2}(1+\delta_{k-1/2,d+1/2})}
                       +\frac{1}{\Delta\tau_{k,d-1/2}(1+\delta_{k-1/2,d-1/2})}}+
      \right.\\\left.
      \frac{v(\nu_k,z_{d-1})}
      {\Delta\tau_{k,d-1/2}(1+\delta_{k-1/2,d-1/2})}\right]\\
=(1+\delta_{k-1/2,d})v(\nu_k,z_{d})-\frac{1}{\Delta\tau_{k,d}}
 \zav{\frac{S(\nu_k,z_{d+1/2})}{1+\delta_{k-1/2,d+1/2}}-
      \frac{S(\nu_k,z_{d-1/2})}{1+\delta_{k-1/2,d-1/2}}}-
\delta_{k-1/2,d}v(\nu_{k-1},z_{d})+\\
\frac{1}{\Delta\tau_{k,d}}\hzav{
 \frac{\delta_{k-1/2,d-1/2}u(\nu_{k-1},z_{d-1/2})}{1+\delta_{k-1/2,d-1/2}}-
 \frac{\delta_{k-1/2,d+1/2}u(\nu_{k-1},z_{d+1/2})}{1+\delta_{k-1/2,d+1/2}}}.
\end{multline}
This system should be supplemented by equations corresponding to the boundary
and initial
conditions. At the outer spatial boundary $z_\text{out}$, we assume no infalling
radiation, consequently $u=v$ and we  derive from Eq.~\eqref{rpzuv2}
\begin{equation}
\frac{1}{\chi(\nu,r)}\frac{\partial v(\nu,p,z)}{\partial z}-
\gamma(\nu,p,z)\frac{\partial v(\nu,p,z)}{\partial \nu}=S(\nu,r)-v(\nu,p,z),
\end{equation}
or, in a difference form
\begin{equation}
\frac{v(\nu_k,z_{d+1})-v(\nu_k,z_{d})}{\Delta\tau_{k,d+1/2}}=
v(\nu_k,z_{d})(1+\delta_{k-1/2,d})-\delta_{k-1/2,d}v(\nu_{k-1},z_{d})-
S(\nu_k,z_{d}).
\end{equation}
The infalling radiation at the inner boundary is taken from the model
atmospheres. The initial solution for $\nu_1$ is derived using the
solution of the radiative transfer equation neglecting the
velocity fields \citep{sphermod,dis}.

The velocity derivatives at the grid points are approximated as in the
hydrodynamical code \citep[see][Eq.~(A.4a) therein]{kki}. The derivatives in the
middle points between grid points are calculated as the average of the
derivatives at the grid points.

The system of algebraic equations Eq.~\eqref{rpzdif} with boundary conditions is
solved using the LAPACK package ({\tt
http://www.cs.colorado.edu/\~{}lapack}, Anderson et al. \citeyear{lapack}).

\end{document}